# Spin Hall effect in a simple classical picture of spin forces

Seng Ghee Tan[1,2,†],    Mansoor B. A. Jalil[2,3]

(1) Data Storage Institute, A*STAR (Agency for Science, Technology and Research), DSI Building, 5 Engineering Drive 1, Singapore 117608

(2) Computational Nanoelectronics and Nano-device Laboratory, Electrical and Computer Engineering Department, National University of Singapore, 4 Engineering Drive 3, Singapore 117576

(3) Information Storage Materials Laboratory, Electrical and Computer Engineering Department, National University of Singapore, 4 Engineering Drive 3, Singapore 117576

Abstract

Spin Hall effect (SHE) in a 2D-Rashba system has been treated in the spin-dependent precession [*J. Sinova et al., Phys. Rev. Letts. 92, 126603 (2004)*.] and the time-space gauge [*T. Fujita et al., New J. Phys. 12, 013016 (2010).*] approaches, both yielding SHE conductivity of $\sigma_y^z = \frac{e}{8\pi}$. Separate studies based on the concept of spin transverse force [*S-Q Shen, Phys. Rev. Lett. 95, 187203 (2005)*.] provide a heuristic but not a quantifiable indication of SHE. In this paper, we provide a more complete description of the SHE using the Heisenberg approach, unifying the Yang-Mills force, the Heisenberg spin force, and the SHE $\sigma_y^z$ under the classical notion of forces and accelerations. Central to this paper is the spin force equations that are satisfied by both $\sigma_y^z$ and $\sigma_x^z$, the Yang-Mills, and the Heisenberg spin forces. By linking $\sigma_y^z$ to the spin forces, one sees that the physics of SHE in a 2D-Rashba system is also a simple classical Lorentz force picture.

## Introduction

Following the first description of the spin Hall effect (SHE) by M.I. Dyakonov and V.I. Perel [1], and an early experimental indication of such effect [2], scientific interest in SHE took a long silence before reemerging only in the 21st century with numerous experimental verifications [3-5] and theoretical discussions [6-8]. SHE was discussed in various systems e.g. the Luttinger, Rashba, cubic and linear Dresselhaus and others with underlying spin orbit coupling. In the Rashba two-dimensional-electron-gas (2DEG) system in particular, SHE was described by Sinova et al. [9] with a theory of electron precession and its transverse separation, leading to a universal SHE conductivity of $\sigma_y^z = e/8\pi$. A few conclusions in this theory seem to also find support in separate studies. For example, the need for a momentum change $\left(\frac{dp_x}{dt}\right)$ means that impurity scattering might destroy SHE. This is consistent with Inoue's [10] prediction that SHE is destroyed by excessive impurities scattering (vertex correction) in a diffusive system. It had also been reasoned that impurities scattering is equivalent to a retardation force which results in average vanishing momentum change $\left(\frac{dp_x}{dt} = 0\right)$ which might in turn destroy SHE. The above collectively suggests that spin Hall in a 2D-Rashba system, with a parabolic energy dispersion can only exist with finite carrier acceleration. Recently, T. Fujita et al. [11,12] used the gauge theoretic approach which makes explicit the role of a derivative magnetic field in time-space [13] on SHE. Assuming the adiabatic relaxation of electron spin to the vector sum of the derivative field (vertical) and the Rashba spin orbit field (in-plane), the approach leads to the same universal $\sigma_y^z = e/8\pi$. Since the special magnetic field required here depends on $\left(\frac{dp_x}{dt}\right)$, once again the crucial role of momentum change (electric field) is necessary. On the other hand, a separate body of work [14-18] which studies the spin-dependent transverse force in terms of classical Yang-Mills field seems to provide a rather heuristic indication of SHE, but not a quantifiable conductivity. The spin transverse force picture has predicted numerous precession correlated electron motion including Zitterbewegung.

We will show in this paper that with the Heisenberg approach, one can provide a broad description of SHE in terms of the spin forces, unifying the Yang-Mills forces, Heisenberg spin forces, and the SHE $\sigma_y^z$ under the classical notion of forces and accelerations. Thus, central to this paper is the derivation of the spin force equations that are satisfied by both $\sigma_y^z$ (spin transverse conductivity) and $\sigma_x^z$ (spin longitudinal conductivity). We use a simple spin orbit gauge in Section A, and a time-space spin orbit gauge in Section B, to derive the equations of motion (EOM) that connect $\sigma_y^z$ and $\sigma_x^z$ to respectively, $f_x^z$ (spin longitudinal force) and $f_y^z$ (spin transverse force). The relations that link $\sigma_y^z$ to $f_x^z$, and likewise $\sigma_x^z$ to $f_y^z$, confirm our idea that the spin transverse forces studied in **Refs.[14-18]** are not responsible for the SHE conductivity. Spin force equations connect SHE conductivities to the Yang-Mills and the Heisenberg spin forces, but SHE conductivities can only be independently derived like in Refs **[9-12].**

### A. Spin Force Equations for a Rashba system

In this section, we will ignore the effect of momentum change $\left(\frac{dp_x}{dt}\right)$, i.e. assuming there is enough scattering to suppress $\left(\frac{dp_x}{dt}\right)$. The spin orbit gauge $A$ used in this section is derived from a simple rewriting of the Hamiltonian. Thus, $A$ in this section is known as the simple spin orbit gauge.

*Classical Yang-Mills*

We will give a brief review of SHE physics described by the classical Yang-Mills fields and the spin transverse forces **[14-18]**. We will begin with the Hamiltonian (vector in bold notation) of a Rashba SOC system in the presence of an external electric field:

$$H = \frac{p^2}{2m} + \lambda\, \boldsymbol{\sigma}.(\boldsymbol{p}\times\boldsymbol{z}) + e\boldsymbol{E_a}.\boldsymbol{r}$$

(1)

where one has $\lambda = \frac{\alpha_R}{\hbar}$, and $\alpha_R$ is the Rashba coupling constant measured in unit of $eVm$, and $E_a$ is the external electric field. It is worth noting that $\lambda$ has the dimension of velocity, i.e. $[\lambda] = m/s$. Thus when written in $H = \gamma\, \boldsymbol{\sigma}.\frac{\lambda}{\gamma}(\boldsymbol{p}\times\boldsymbol{z})$, one has an effective magnetic field of $\boldsymbol{b} = \frac{\lambda}{\gamma}(\boldsymbol{p}\times\boldsymbol{z})$ where $\boldsymbol{b}$ has the dimension $[b] = Tesla$, and $[\gamma] = Joule/Tesla$. In the language of gauge theory, the Rashba SOC energy is equivalent to the presence of the simple spin orbit gauge $\boldsymbol{A} = \frac{m\lambda}{e}(\sigma_y, -\sigma_x, 0)$, so that up to the first order in SOC constant, the Hamiltonian now reads $H = \frac{1}{2m}(\boldsymbol{p}-e\boldsymbol{A})^2 + e\boldsymbol{E_a}.\boldsymbol{r}$. The curvature of the gauge potential has the physical meaning of the classical Yang-Mills magnetic or electric fields as given by $F_{\mu\nu} = \frac{ie}{\hbar}[A_\mu, A_\nu]$. The $z$ component of this effective field will be magnetic,

$$B_z \boldsymbol{n_z} = F_{xy}\boldsymbol{n_z} = -\frac{2m^2\lambda^2}{e\hbar}\sigma_z \boldsymbol{n_z}$$

(2)

in which $\boldsymbol{n_z}$ is a vertical unit vector. Due to the $\sigma_z$ coefficient, this effective Yang-Mills magnetic field can be treated like it is spin-dependent, i.e. it points up (down) when "seen" by electron of spin down (up).

This field, coupled with classical electron motion, thus generates Lorentz-like force acting on the moving electrons. For example, if a spin-down electron initially moving along $x$-direction, it will feel a transverse force and move to the left, likewise, spin-up will experience a force in the opposite direction and move to the right, leading to the separation of spins and hence

$$\boldsymbol{f^{YM}} = eB_z\boldsymbol{v} \times \boldsymbol{n_z}$$

(3)

The above is a "force" expression in the classical sense of Lorentz force and Hall effects. One can thus derive the Yang-Mills force as follows

$$f_k^{YM} = \frac{ie^2}{m\hbar}(P_\mu + eA_\mu)[A_k, A_\mu]$$

(4)

The physics is pictorially illustrated in Fig.1 which shows a Lorentz-force effect on electron of one spin orientation. The spin transverse force is denoted by $f_y$.

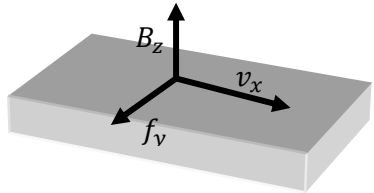

Fig.1. The picture of spin transverse force experienced by electron traveling along x at a constant velocity.

The above indeed gives a heuristic, and intuitive picture of spin separation due to $f_y$, but the actual dynamics of electron remains unclear. For example, spin undergoes precession about the spin-orbit effective magnetic field **b** (not the Yang-Mills **B** of Eq.2). As a result of this imagination, the Yang-Mills **B** field acting on each electron also flips up or down at the rate of precession, leading to the transverse oscillation of $f_y$. Thus spin transverse force leads to spin transverse oscillation rather than the permanent spin separation of SHE.

*Heisenberg approach*

We will now show using the Heisenberg approach that the spin orbit gauge could lead to a set of spin force equations which relate electron acceleration due to $\frac{dp_x}{dt}$ to spin conductivity of $\sigma_y^z$ and $\sigma_x^z$. We start with the classical equation of motion for force i.e. $f = m\frac{dv}{dt}$. Since in classical mechanics, $= \frac{\partial H}{\partial p}$, one can rewrite the force equation but now taking $v$ as an operator

$$f = m\frac{d\langle v \rangle}{dt} = \frac{d\langle \boldsymbol{p} - e\boldsymbol{A} \rangle}{dt}$$

(5)

Applying the quantum dynamic of $i\hbar \frac{d\langle O \rangle}{dt} = \langle [\,O, H\,] \rangle + i\hbar \langle \frac{\partial O}{\partial t} \rangle$ to the velocity operator, one arrives at

$$f = \left(\frac{1}{i\hbar}\langle[\boldsymbol{p},H]\rangle + \langle\frac{\partial \boldsymbol{p}}{\partial t}\rangle\right) + \left(-\frac{e}{i\hbar}\langle[\,\boldsymbol{A},H\,]\rangle - e\langle\frac{\partial \boldsymbol{A}}{\partial t}\rangle\right) \equiv \boldsymbol{f_1} + \boldsymbol{f_2}$$

(6)

where $\boldsymbol{A} = G(\boldsymbol{\sigma} \times \boldsymbol{E_R})$, one has $\langle\frac{\partial \boldsymbol{A}}{\partial t}\rangle = G\langle \boldsymbol{\sigma} \times \frac{\partial \boldsymbol{E}}{\partial t} + \frac{\partial \boldsymbol{\sigma}}{\partial t} \times \boldsymbol{E_R}\rangle$ and $\boldsymbol{E_R}$ is the electric field associated with the Rashba effect. For physical clarity, we note that $G = \frac{m\lambda}{eE_R}$ would have the dimension of $[G] = time$, since $[\lambda] = velocity$. Thus,

$$\boldsymbol{f_1} = \frac{1}{i\hbar}\langle\left[\boldsymbol{p},\left(\frac{p^2}{2m} + \frac{e(\boldsymbol{p}.\boldsymbol{A} + \boldsymbol{A}.\boldsymbol{p})}{2m} + e\,\boldsymbol{E}.\boldsymbol{r}\right)\right]\rangle = \frac{e}{i\hbar}\langle\left[\boldsymbol{p},\frac{(\boldsymbol{p}.\boldsymbol{A} + \boldsymbol{A}.\boldsymbol{p})}{2m} + \boldsymbol{E}.\boldsymbol{r}\right]\rangle$$

(7a)

$$\boldsymbol{f_2} = \frac{-e}{i\hbar}\langle\left[\boldsymbol{A},\left(\frac{p^2}{2m} + \frac{e(\boldsymbol{p}.\boldsymbol{A} + \boldsymbol{A}.\boldsymbol{p})}{2m} + e\,\boldsymbol{E}.\boldsymbol{r}\right)\right]\rangle = \frac{ie}{\hbar}\langle\left[\boldsymbol{A},\frac{e(\boldsymbol{p}.\boldsymbol{A} + \boldsymbol{A}.\boldsymbol{p})}{2m}\right]\rangle$$

(7b)

For consistency, the longitudinal direction is always taken to be $x$, the transverse direction $y$. As the term $\left[\boldsymbol{p},\frac{(\boldsymbol{p}.\boldsymbol{A}+\boldsymbol{A}.\boldsymbol{p})}{2m}\right]$ vanishes when the spin orbit constant $G$ is spatially uniform, which is assumed in this paper to simplify analysis, $\boldsymbol{f_1}$ is reduced to pure electrical force, i.e. $\boldsymbol{f_1} = e\boldsymbol{E}$. This is true regardless of the quantum state of the carrier because $\boldsymbol{E}$ is not an operator. Thus, referring back to Eq.(5) one could see that $\frac{d\langle p \rangle}{dt} = e\boldsymbol{E}$, which physically means that external electrical force has a direct effect on the electron's momentum but not necessarily its velocity. Although in 2DEG, or metallic system, velocity is a linear function of the momentum, this would not be the case in other non-parabolic energy system. For a general spin orbit gauge $A$, one can obtain $f_2 = \frac{ie}{\hbar}\langle [A, \frac{e(p.A+A.p)}{2m}]\rangle$, a general expression of the Heisenberg spin force is

$$f_{2i} = \frac{ie^2}{m\hbar}\langle P_\mu [A_i, A_\mu]\rangle$$

(8)

which resembles the classical Yang-Mills force of Eq.(4). Combining $f_1$ and $f_2$, and ignoring the higher order terms, the spin force equations can be written as

$$m\frac{d\langle v_i\rangle}{dt} = eE_i + f_i^{YM}$$

(9)

Thus, the heuristic Yang-Mills force has been fitted by comparisons with the Heisenberg spin force, and with approximations, into the spin force equation, providing a clearer picture of how these spin forces generate electron spin motion. One can of course be contented that a similar spin force equation can be derived using just Eq.(4) in the Yang-Mills picture, and in purely classical sense. But the Heisenberg approach provides a quantum mechanical verification to the physics. Applying Heisenberg spin force $f_2$ to the simple spin orbit gauge $A = \frac{m\lambda}{e}(\sigma_y, -\sigma_x, 0)$, one produces

$$f_2 = \frac{m\lambda^2}{\hbar}\langle\{\sigma_z, p_y\}i - \{\sigma_z, p_x\}j\rangle$$

(10)

Of physical importance is the fact that each force component is perpendicular to a momentum coupled to it. For example, the term $\{\sigma_z, p_y\}i$ contains $p_y$. Combining $f_1$ and $f_2$, the spin force equations are:

$$m\frac{d\langle v_x\rangle}{dt} = eE_x + \frac{m\lambda^2}{\hbar}\langle\{\sigma_z, p_y\}\rangle$$

(11a)

$$m\frac{d\langle v_y\rangle}{dt} = eE_y - \frac{m\lambda^2}{\hbar}\langle\{\sigma_z, p_x\}\rangle$$

(11b)

*SHE conductivity*

We have seen that the Yang-Mills forces have been related to the Heisenberg spin forces. We will now set out to study their relations to the SHE conductivity. The spin current operator, that has been used in **Refs [9,11,12]** to derive SHE current or conductivity is:

$$j^z = \frac{\hbar}{4}\{v, \sigma_z\} = \frac{\hbar}{4m}\{(p - eA), \sigma_z\},$$

(12)

where spin current has the dimension of $[j^z] = angular\ momentum/second$. One notes that simple $A = \frac{m\lambda}{e}(\sigma_y, -\sigma_x, 0)$ vanishes from (12) by virtue of $\{A, \sigma_z\} = 0$, thus resulting in $j^z = \frac{\hbar}{4m}\{p, \sigma_z\}$ which would prompt one to deduce that simple $A$ is not directly related to spin conductivity $\sigma_y^z$. In fact Rashba spin orbit coupling is only needed to generate the required vertical spin polarization. Comparing 11(a) & (b) with $j^z = \frac{\hbar}{4m}\{p, \sigma_z\}$, one has

$$\begin{pmatrix} j_x^z \\ j_y^z \end{pmatrix} = \frac{\hbar^2}{4m^2\lambda^2}\begin{pmatrix} f_{2y} \\ f_{2x} \end{pmatrix},$$

(13)

With this, the SHE conductivity/current has been connected to the Yang-Mills and the Heisenberg spin forces. As an aside, one notes by inspection that the SHE conductivity (current) $\sigma_y^z(j_y^z)$ is related to the spin longitudinal force, but not the spin transverse force. One can then modify Fig.1. and construct a new picture below

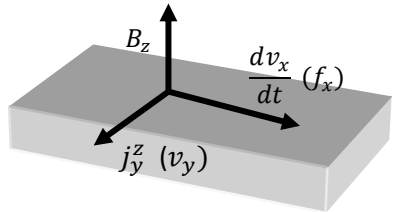

Fig.2. A modified Lorentz force picture which describes SHE current as related to the spin longitudinal force.

Compared to Fig.1., the modified Lorentz force picture is a rotated version (about $z$ axis over an azimuthal angle) of Fig.1. Finally, one writes the spin force equation that unify the Yang-Mills forces, Heisenberg spin forces, and SHE conductivity

$$m\frac{d\langle v_i\rangle}{dt} = eE_i + \frac{4m^2\lambda^2}{\hbar^2}\langle\psi_{ip}|j_j^z|\psi_{ip}\rangle\varepsilon_{ijz}$$

(14)

Note that $|\psi_{ip}\rangle$ is the in-plane eigenstate of the Rashba 2D system, or the spin eigensolution of Eq.(1). As mentioned earlier that once the link is established through the above, one can proceed to independently derive the explicit expression of $\sigma_y^z$ or $j_y^z$, noting that $j_y^z = \sigma_y^z E_x$. The explicit expression of the SHE conductivity (current) $\sigma_y^z(j_y^z)$ will depend on the quantum state of the carrier i.e. $|\psi_{ip}\rangle$, which can be obtained by solving the Hamiltonian. Figure 3 gives a pictorial summary of the theoretical processes in section A.

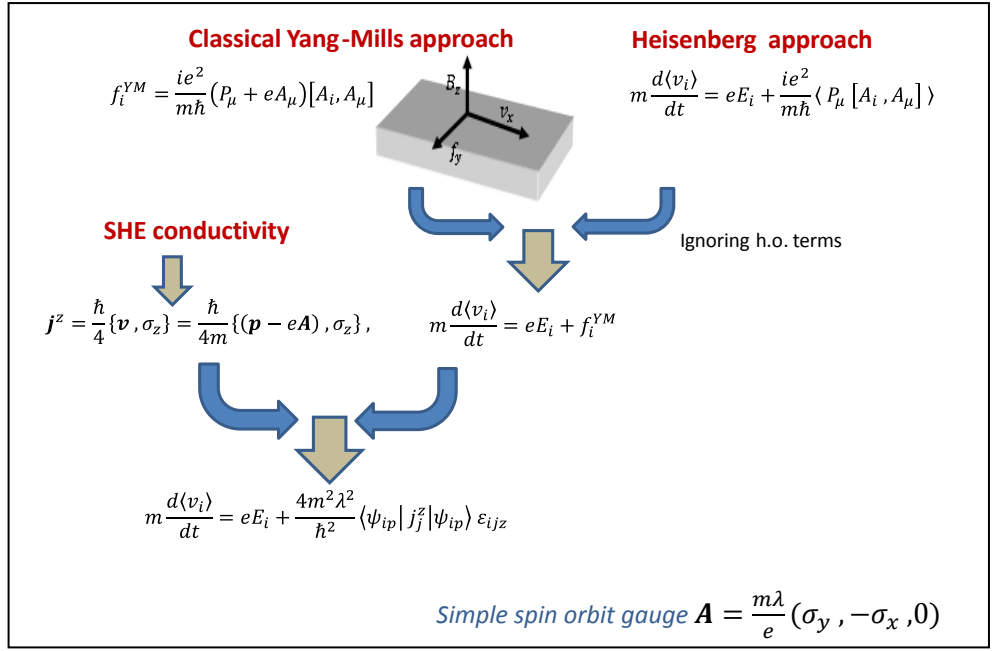

Fig.3. Unification of Yang-Mills force, Heisenberg spin force, SHE conductivity under the Heisenberg spin force equation. The simple spin orbit gauge is used in the above.

### B. Spin Force Equations for a Rashba System with Time-gauge

This section revisits the theoretical processes in Section A except the spin orbit gauge used here is the time-space version, which describes a Rashba system with finite momentum change $(\frac{dp_x}{dt})$. As a matter of fact, with the simple spin orbit gauge of Section A, SHE conductivity (current) $\sigma_y^z(j_y^z)$ as shown in Eq.(14), $\langle \psi_{ip}(t)|j_y^z(t)|\psi_{ip}(t)\rangle$ vanishes because the spin state of electron $|\psi_{ip}(t)\rangle$ in Rashba 2DEG is generally in-plane, while the spin current operator contains a Pauli-$z$ matrix. Although we know that in reality, spin tilts out-of-plane as electron experiences a momentum change of $(\frac{dp_x}{dt})$ **[11, 12]**, this effect has not been reflected in the Hamiltonian of Section A. The Hamiltonian used in Eq.(1) is the time-independent form ($H_s$). One could check that the Heisenberg SHE current $j_y^z(t) = e^{iH_s t} j_y^z e^{-iH_s t}$ evolves with the generator $H_S$. But this will not produce the effect of momentum change or electric curvature. On the other hand, the eigenstate of $H_S$ which is $|\psi_{ip}(t)\rangle$ will only evolve in dynamic phase. Therefore, to obtain non-vanishing SHE conductivity (current) $\sigma_y^z(j_y^z)$, one needs the time-dependent treatment **[11, 12]**, where an electric curvature (effective magnetic field) appears vertical to the 2D plane via a local gauge transformation in the momentum space-time. A sum field ($\mathbf{\Omega}$) which is the vector sum of the vertical time-field $\frac{\hbar(\partial_t \mathbf{b} \times \mathbf{b})}{\gamma b^2}$ and the in-plane spin orbit field ($\mathbf{b}$) could thus be defined. Electron with spin aligned parallel/antiparallel to the total field is identified with the $\rho = \pm$ bands. In summary, one ventures from the original $H_S$ to a time-dependent $H_I(t)$, and performs the local gauge transformation. An inverse rotation would yield the Hamiltonian with the electric curvature. In the accelerating frame $(\frac{dp_x}{dt})$ of electron, one obtains back a time-independent Hamiltonian of

$$H_s^f = \frac{p^2}{2m} + \gamma\, \boldsymbol{\sigma}.\boldsymbol{b} + \hbar \boldsymbol{\sigma}.\boldsymbol{a} + e\boldsymbol{E}.\boldsymbol{r}$$

(15)

where $\boldsymbol{a}$ is perpendicular to $\boldsymbol{b}$, and the explicit expression is $\hbar \boldsymbol{a} = \left(\frac{\hbar(\partial_t \boldsymbol{b} \times \boldsymbol{b})}{b^2}\right) = \frac{\hbar e E_x p_y}{p^2}\boldsymbol{z}$. We note in passing that two transformations have been performed to arrive at Eq.(15) see Ref **[11, 12]** for details. In fact, noting that $\boldsymbol{b} = \frac{\lambda}{\gamma}(\boldsymbol{p} \times \boldsymbol{z})$, one could reason with classical physics that as spin aligns to the sum field $\mathbf{\Omega}$, a vertical component of spin polarization $\langle s \rangle = \rho s\, \Omega/\Omega$ is generated, $s$ is the spin quantum number. Now, it can be derived, noting that $= \alpha_R/\hbar$, and using the time-independent $H_s^f$ that the time-space spin orbit gauge is:

$$\boldsymbol{A} = \frac{m\lambda}{e}(\sigma_y, -\sigma_x + \frac{\hbar e E_x}{p^2 \lambda}\sigma_z, 0)$$

(16)

Note that the momentum $p$ is in-plane. With Eq.7(b), the Heisenberg spin force equations for the time-space spin orbit gauge is

$$m\frac{d\langle v_x \rangle}{dt} = eE_x + \frac{m\lambda^2}{\hbar}\langle\psi_\rho|\{\sigma_z, p_y\}|\psi_\rho\rangle + \frac{eE_x m\lambda}{p^2}\langle\psi_\rho|\{\sigma_x, p_y\}|\psi_\rho\rangle$$

$$= eE_x + \frac{4m^2\lambda^2}{\hbar^2}\langle\psi_\rho|j_y^z|\psi_\rho\rangle + \frac{eE_x 4m^2\lambda}{\hbar p^2}\langle\psi_\rho|j_y^x|\psi_\rho\rangle$$

(17a)

$$m\frac{d\langle v_y\rangle}{dt} = eE_y - \frac{m\lambda^2}{\hbar}\langle\psi_\rho|\{\sigma_z,p_x\}|\psi_\rho\rangle - \frac{eE_y m\lambda}{p^2}\langle\psi_\rho|\{\sigma_x,p_x\}|\psi_\rho\rangle$$

$$= eE_y - \frac{4m^2\lambda^2}{\hbar^2}\langle\psi_\rho|j_x^z|\psi_\rho\rangle - \frac{eE_y 4m^2\lambda}{\hbar p^2}\langle\psi_\rho|j_x^x|\psi_\rho\rangle$$

(17b)

The SHE conductivity (current) $\sigma_y^z(j_y^z)$ of $j_\mu^z = \frac{\hbar}{4}\langle\{v_\mu,\sigma_z\}\rangle = \frac{\hbar}{4m}\langle\psi_\rho|\{(p_\mu - eA_\mu),\sigma_z\}|\psi_\rho\rangle$ can, once again, be fitted with approximations into the spin force equations. Different from the simple Rashba system, in the time-space gauge, the expression $\frac{m\lambda^2}{\hbar}\langle\psi_\rho|\{\sigma_z,p_y\}|\psi_\rho\rangle$ is non-vanishing. In fact it can be derived from the polarization expression of $\langle s \rangle = \rho s\, \Omega/\Omega$, which is non-vanishing, and finite. Therefore, the SHE conductivity (current) $\sigma_y^z(j_y^z)$ will be non-vanishing consistent with previous accounts. Figure 4 summarizes the theoretical processes similar to those in Section A except the above is carried out using the time-space spin orbit gauge, instead of the simple spin orbit gauge.

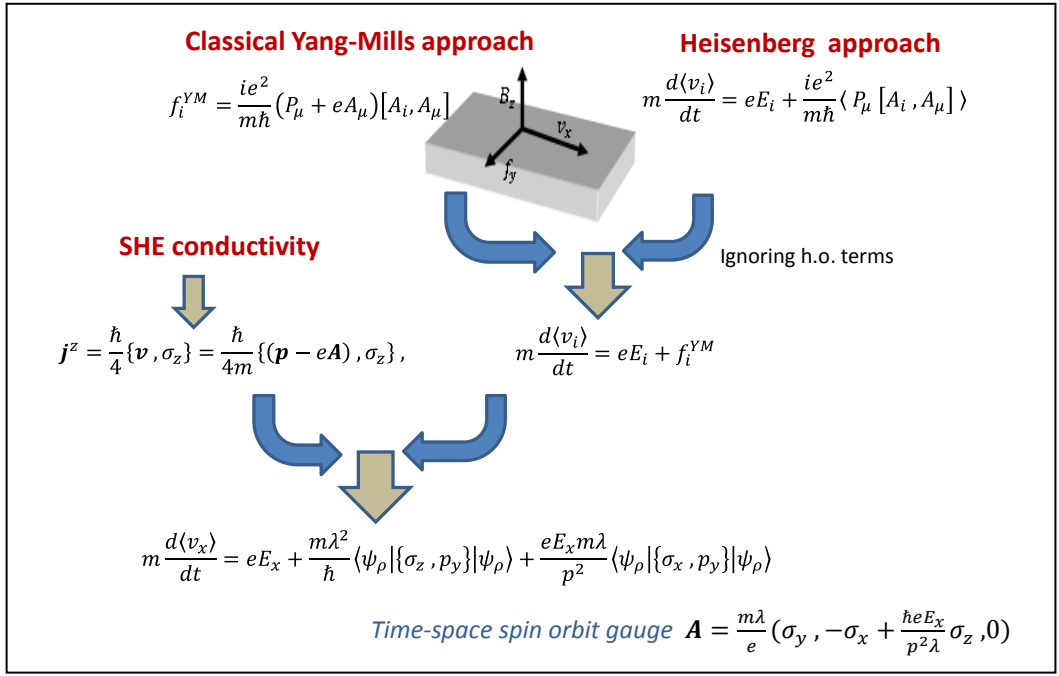

Fig.4. Unification of Yang-Mills force, Heisenberg spin force, SHE conductivity under the Heisenberg spin force equation. The time-space spin orbit gauge is used in the above.

**Conclusion**

The Yang-Mills forces, Heisenberg spin forces, and SHE conductivities have been derived and linked under the Heisenberg spin force equations. While fitting SHE $\sigma_y^z$ or $j_y^z$ into the Heisenberg spin force equations, an intuitive Lorentz-force picture description of the SHE is observed. Since explicit quantum mechanical expression of $\sigma_y^z$ or $j_y^z$ can be derived independently, our work has no bearing on previous derivation of SHE $\sigma_y^z$ except we have found their relationship to the spin forces. Both the simple spin orbit gauge and the time-space spin orbit gauge have been studied.